\documentclass[aps,prx,reprint,groupedaddress,twocolumn,longbibliography]{revtex4-2}

\usepackage{amsmath}
\usepackage{graphicx}
\usepackage{upgreek}
\usepackage{graphicx}
\usepackage{comment}
\usepackage{appendix}

\usepackage{hyperref}

\usepackage{xcolor}

\newcommand{\trap}{\mathrm{trap}}

\newcommand{\D}{\mathrm{D}}
\newcommand{\W}{\mathrm{W}}
\newcommand{\ts}{\thinspace}

\begin{document}

\title{Self-Adaptive Stabilization and Quality Boost for Electron Beams \\ from All-Optical Plasma Wakefield Accelerators}

\author{
D. Campbell$^{1,2,4*}$,
T. Heinemann$^{1,3,4\dagger}$,
A. Dickson$^{1,4,5}$,
T. Wilson$^{1,4,9}$,
L. Berman$^{1,4}$,
M. Cerchez$^{3}$,
S. Corde$^{6}$,
A. Döpp$^{2}$,
A.F. Habib$^{1,4}$,
A. Irman$^{7}$,
S. Karsch$^{2}$,
A. Martinez de la Ossa$^{8}$,
A. Pukhov$^{9}$,
L. Reichwein$^{5,9}$,
U. Schramm$^{7}$,
A. Sutherland$^{1,3,4}$,
B. Hidding$^{1,3,4\ddagger}$
}

\affiliation{
$^{1}$Department of Physics, SUPA, University of Strathclyde, Glasgow, UK \\
$^{2}$Ludwig-Maximilians-Universität München, Am Coulombwall 1, 85748 Garching, Germany \\
$^{3}$Institute for Laser and Plasma Physics, Heinrich Heine University Düsseldorf, 40225 Düsseldorf, Germany \\
$^{4}$Cockcroft Institute, Sci-Tech Daresbury, Keckwick Lane, Daresbury, Cheshire WA4 4AD, UK \\
$^{5}$Peter Grünberg Institut (PGI-6), Forschungszentrum Jülich, 52425 Jülich, Germany \\
$^{6}$LOA, ENSTA, CNRS, École Polytechnique, Institut Polytechnique de Paris, 91762 Palaiseau, France \\
$^{7}$Helmholtz-Zentrum Dresden–Rossendorf, Institute of Radiation Physics, Bautzner Landstraße 400, 01328 Dresden, Germany \\
$^{8}$Deutsches Elektronen-Synchrotron DESY, 22607 Hamburg, Germany \\
$^{9}$Institute for Theoretical Physics I, Heinrich Heine University Düsseldorf, 40225 Düsseldorf, Germany
}

\thanks{david.j.campbell@strath.ac.uk\\$^\dagger$thomas.heinemann@hhu.de\\$^\ddagger$bernhard.hidding@hhu.de}

\date{\today}

\begin{abstract}

Shot-to-shot fluctuations in electron beams from laser wakefield accelerators present 
a significant
challenge for applications.
Here, we show that instead of using such fluctuating beams directly, employing them to drive a plasma photocathode-based wakefield refinement stage can produce secondary electron beams with greater stability, higher quality, and improved reliability. 
Our simulation-based analysis reveals that drive beam jitters are compensated by both the insensitivity of beam-driven plasma wakefield acceleration, and the decoupled physics of plasma photocathode injection.  
While beam-driven, dephasing-free plasma wakefield acceleration mitigates energy and energy spread fluctuations, intrinsically synchronized plasma photocathode injection compensates charge and current jitters of incoming electron beams, and provides a simultaneous quality boost.   
Our findings suggest plasma photocathodes are ideal injectors for hybrid laser-plasma wakefield accelerators, and 
nurture prospects for demanding applications such as free-electron lasers.
\end{abstract}


\maketitle

\section{INTRODUCTION}

Laser wakefield acceleration (LWFA)~\cite{Tajima1979,Malka2002a,ManglesNature2004,GeddesNature2004,Faure2004} utilizes a laser pulse with an intensity \( I_L \gtrsim 10^{18}\,\mathrm{W/cm^2} \) to expel plasma electrons transversely, thereby driving a collective plasma wave in its wake.
This resulting plasma structure supports longitudinal electric fields exceeding several hundred gigavolts per meter (GV/m), enabling rapid 
electron acceleration over short distances.

Electrons can be captured into the wake and subsequently form an injected electron `witness' beam. This injection can be achieved through various processes, ranging from self-injection~\cite{ManglesNature2004,GeddesNature2004,Faure2004} to more elaborate methods involving local modifications of plasma density~\cite{Bulanov1998Downramp,gonsalves2011NatureTunableDensity,Buck2013Downramp,Picksely2023-channel-density-injection}, multiple intense laser pulses~\cite{Faure2006}, using higher ionization levels and dopants~\cite{ChenJAP200610short,ZengMing-2014-STII-Self-truncated-PoP,couperus2017demonstration,Irman2018}, as well as combinations of these methods~\cite{thaury2015shock,Wenz2019}. 

A common feature across these schemes is that electron injection is tightly coupled to the wake excitation process.
Fluctuations in the drive laser’s characteristics -- such as power, polarization~\cite{Ma.Y-2020-Polarization-Injection-LWFA}, or wavefront quality~\cite{Popp2010PhysRevLett.105.215001} -- along with the inherently nonlinear and relativistic nature of the laser–plasma interaction~\cite{esareyOverview}, significantly influence the evolution of the plasma wakefield.
These coupled dynamics, in turn, affect injection rates and charge yield, resulting in shot-to-shot variations in the properties of the injected electron beam.

Once injected, the LWFA-inherent challenges of diffraction of the laser driver and dephasing with the injected electrons are further complications that can change the longitudinal and transverse phase space of the injected electron beam throughout the subsequent acceleration process~\cite{esareyOverview}. 
While meticulous control and fine-tuning of the laser-plasma interaction can improve stability, state-of-the-art systems still exhibit fluctuations in charge, energy, and energy spread on the order of a few to 10\% shot-to-shot~\cite{Wenz2019,MaierPRX,Shalloo2020-LWFA-ML-Automation}.

However, engineering optimization of LWFA is not the only path toward improved stability.
Here, we explore intrinsic compensation mechanisms when using fluctuating LWFA output to power a subsequent electron beam-driven plasma wakefield acceleration (PWFA)~\cite{Chen1985PhysRevLettshort,Rosenzweig1988PRLshort} refinement stage~\cite{HiddingPRL2010PhysRevLett.104.195002short,Hidding2014,Hidding2019FundamentalsLWFA-PWFA,Alberto-Hybrid,Gilljohann2019Hybrid,KurzNatComm2021,Couperus2021PWFADensityDownramp}.
In PWFA, strong plasma waves are excited by the unipolar Coulomb force of an electron beam, but with peak driver fields orders of magnitude lower than in LWFA.
This environment enables controlled electron injection via a plasma photocathode~\cite{Hidding2012, YunfengPhysRevSTABshort, Deng2019Trojan}, where a low-power laser pulse with $I_L \sim 10^{15}-10^{16}\,\text{W}/\text{cm}^2$, just above the tunneling ionization threshold, releases electrons directly into the wake, with negligible momentum imparted by the injection laser~\cite{Habibplasma-photocathodes-2023}.

The plasma photocathode can thus generate well-defined electron bunches with potentially ultralow normalized emittance $\epsilon_\mathrm{n}<100\,\text{nm}\,\text{rad}$ root mean square (r.m.s.)~\cite{Hidding2012,YunfengPhysRevSTABshort,schroeder2014PRSTABTrojan,Deng2019Trojan,Habibplasma-photocathodes-2023}, while the release process and injection rates are decoupled from the driver and wakefield configuration.
While staging of plasma wakefield accelerators is usually associated with challenges in capture efficiency and beam quality preservation \cite{Steinke2016MultistageAccelerators}, here
we show that staged electron beam utilization from LWFA into plasma photocathode-based PWFA can yield electron output bunches with significantly higher stability, in addition to substantially enhanced bunch quality.

To explore the interplay of key mechanisms in this stabilization process, we perform high-fidelity, quasi-3D particle-in-cell simulations of the PWFA refinement stage equipped with a plasma photocathode using FBPIC~\cite{LEHE-FBPIC-1}.
The electron driver beams utilized in the study encompass a wide parameter space accessible by
contemporary LWFA systems via a variety of methods, including, but not limited to,
self-truncated ionization injection~\cite{ZengMing-2014-STII-Self-truncated-PoP,couperus2017demonstration, Irman2018} or shock-front injection~\cite{Buck2013Downramp,GoetzfriedPRX2020, HUE-Density-Downramp,Moritz2022-PRX,v.Grafenstein2023-GeV-Nanocoloumb}.

\begin{figure}[!htbp]
\includegraphics[width=0.45\textwidth]{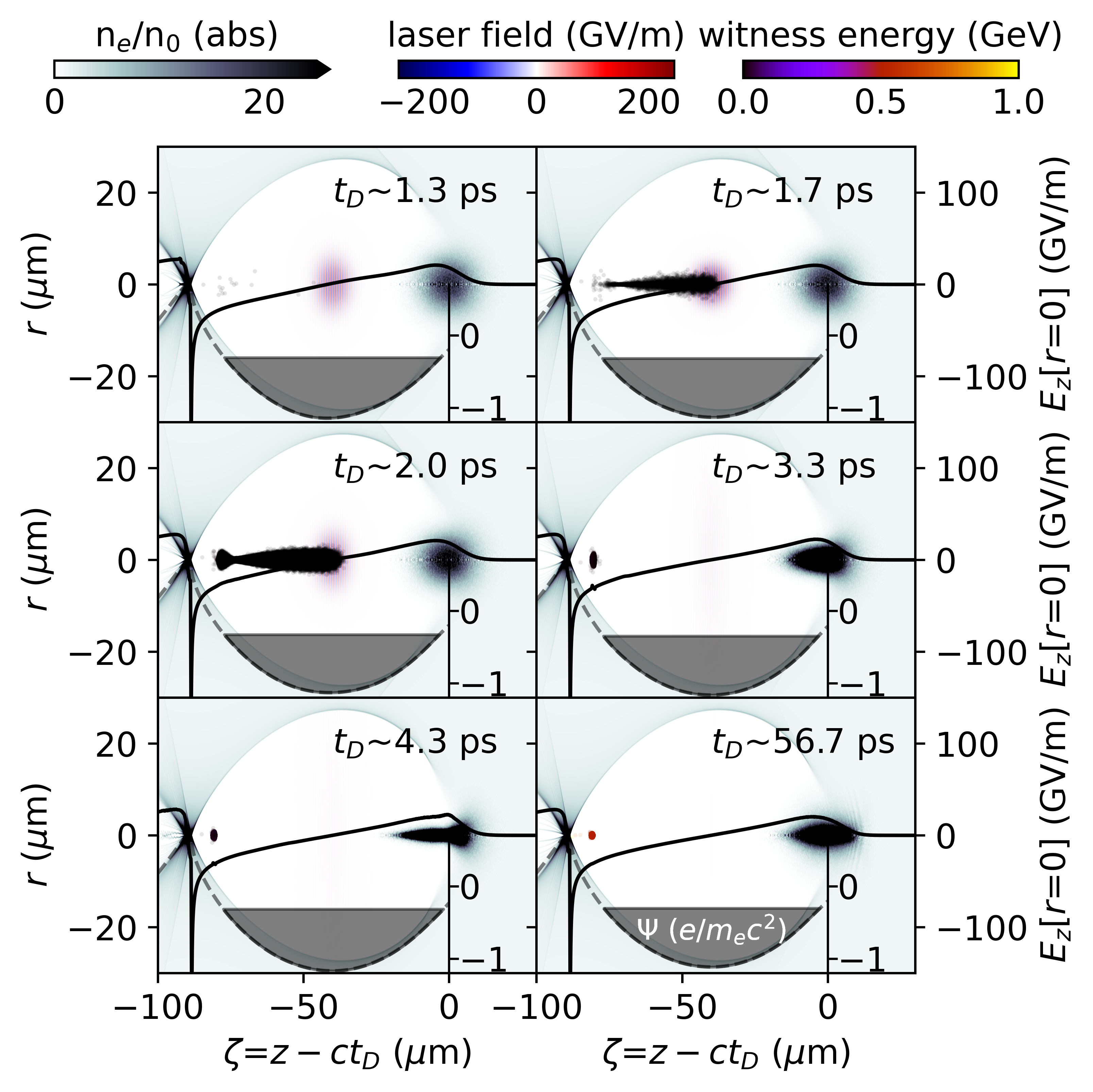}
\caption{PWFA showcase simulation:
Simulation snapshots illustrate plasma photocathode injection within a helium plasma wave. The electrostatic pseudo-potential $\Psi$ is shown as a grey dashed line, and the shaded region indicates the longitudinal trapping window ($\Psi - \Psi_{\mathrm{max}} \leq -m_\mathrm{e}c^2/e$) within the blowout structure. The corresponding on-axis electric field $E_z$ is plotted as a solid black line. An electron beam drives the wakefield as it propagates to the right. A co-propagating laser pulse (blue to red) ionizes He$^{2+}$, releasing witness electrons that are color-coded by energy (black to yellow).
}
\label{set-up}
\end{figure}
For the showcase PWFA configuration, the plasma density is set to \( n_\mathrm{e} \approx 1.12 \times 10^{17}\,\mathrm{cm}^{-3} \), corresponding to a cold plasma wavelength \(\lambda_\mathrm{p} = 2\pi c / \omega_\mathrm{p} \approx 100\,\upmu\mathrm{m}\), where the plasma frequency is defined as \(\omega_\mathrm{p} = \sqrt{n_\mathrm{e} e^{2} / (m_{\mathrm{e}} \epsilon_0)}\) and is related to the plasma wave number via \(\omega_\mathrm{p} = c k_\mathrm{p}\).

The baseline electron driver beam is bi-Gaussian with charge \( Q_\mathrm{D} = 500\,\mathrm{pC} \), length \(\sigma_z = 6.0\,\upmu\mathrm{m}\) (r.m.s.), radius \(\sigma_r = 4.0\,\upmu\mathrm{m}\) (r.m.s.), mean kinetic energy \( W_\mathrm{D} = (\gamma_\mathrm{D} - 1) m_\mathrm{e} c^{2} = 500\,\mathrm{MeV} \), where \(\gamma_\mathrm{D}\) is the beam’s Lorentz factor, relative energy spread \(\Delta W_\mathrm{D} / W_\mathrm{D} = 10\%\) (r.m.s.), $\epsilon_\mathrm{n} = 2.0~\upmu\mathrm{m\, rad}$ (r.m.s.), and peak current \( I_\mathrm{p} \approx 10\,\mathrm{kA} \).

By choosing these values, neither the longitudinal resonance condition \cite{Katsouleas1986} \(\sigma_{z,\mathrm{m}} = \sqrt{2} / k_\mathrm{p} \approx 22.5\,\upmu\mathrm{m}\)  nor the transverse matching condition \cite{Lu2006} \(\sigma_{r,\mathrm{m}} = \sqrt[4]{2 / \gamma_\mathrm{D}} \sqrt{\epsilon_\mathrm{n} / k_\mathrm{p}} \approx 1\,\upmu\mathrm{m}\) is fulfilled.
This is intentional, to account for non-ideal conditions that may arise from unoptimized beam transport between stages -- further complicated by the inherent volatility of the LWFA electron beam output -- and to underscore the robustness of the plasma photocathode-enhanced PWFA stage.
This choice is also appropriate because, in present day hybrid LWFA$\rightarrow$PWFA \cite{Hidding_2023_P_ProgressHybridPlasma} setups, the electron beams emerging from the LWFA stage are typically intrinsically short, carry multi-kA currents  \cite{Lundh2011,Heigoldt2015,Zarini2022-CTR-LWFAPRAB}, 
and expand rapidly in the transverse dimension between stages.

Nevertheless, the driver beam's peak density, expressed as  
\( n_\mathrm{b} = Q_\mathrm{D} / [e(2\pi)^{3/2} \sigma_r^2 \sigma_z] \approx 2.1 \times 10^{18}\,\mathrm{cm}^{-3} \), exceeds \( n_\mathrm{e} \) by more than an order of magnitude.
The driver is therefore sufficiently 
overdense to enter the nonlinear blowout regime of PWFA~\cite{RosenzweigPRA1991}, even under substantial driver charge variations.
The blowout regime is characterized by longitudinally uniform linear focusing fields and a radially uniform accelerating field with a linear slope around the wake center.
This linear field configuration is a prerequisite for preserving the normalized emittance during acceleration~\cite{RosenzweigPRA1991}.

Utilizing intrinsically short driver beams from LWFA, confined to the front of the blowout, is furthermore beneficial for plasma photocathodes, as this eliminates any influence of the driver fields on the plasma photocathode laser field ionization rate, as well as transverse momentum and witness beam emittance growth, over a wide range of possible release positions~\cite{YunfengPhysRevSTABshort,Deng2019Trojan,Habibplasma-photocathodes-2023}. 
The radial electric field of the driver beam is given by  
\( E_r(r) = (Q_\mathrm{D} / [(2\pi)^{3/2} \sigma_z \epsilon_0 r]) \times \left(1 - \exp(-r^2 / 2\sigma_r^2)\right) \),  
and peaks at approximately \( 67\,\mathrm{GV/m} \).
This is sufficient to field-ionize helium from its ground state (low ionization threshold, LIT: 24.6\,eV), while leaving its second, higher ionization threshold level (HIT: 54.4\,eV) unperturbed.
In the present study, we assume preionization of the first helium level using an auxiliary laser pulse, as recently demonstrated experimentally~\cite{Gilljohann2019Hybrid, KurzNatComm2021, Couperus2021PWFADensityDownramp, Moritz2022-PRX}.
This controlled preionization supports a more stable and robust wakefield formation~\cite{Gilljohann2019Hybrid, KurzNatComm2021}.

This hybrid LWFA$\rightarrow$PWFA configuration is then equipped with a collinear plasma-photocathode laser, delayed by approximately 133~fs with respect to the driver centroid. The laser operates with an energy of \( W_L = 0.1\,\mathrm{mJ} \) and a pulse duration of \( \tau = 30\,\mathrm{fs} \) full width at half maximum (FWHM). 
It is focused to an r.m.s. spot size of \( w_0 = 5\,\upmu\mathrm{m} \) at the blowout center -- a unique feature of the plasma photocathode that gives rise to several inherent stability mechanisms discussed in this paper. 
This setup enables witness bunch generation by controlled ionization of the HIT (He$^+$) ions.

Figure \ref{set-up} illustrates the showcase PWFA simulation configuration, showing the plasma photocathode process as the injector laser traverses its focus.
Consecutive PIC simulation snapshots are shown in the co-moving coordinate system \(\zeta = z - c t_\D\), given that the wake propagates at approximately \(c\) due to the relativistic velocity of the driver. Here, \( t_\D \) denotes the driver’s propagation time in the plasma.

The electrostatic pseudo-potential \(\Psi\) and the corresponding longitudinal component of the on-axis electric field (\(E_z=\partial_\zeta\Psi\)) are shown in Fig.~\ref{set-up}, with the zero crossing located at approximately \(\zeta \approx -40\,\upmu\mathrm{m}\).
Electrons are released at rest by the plasma photocathode laser, with the laser delay chosen such that the ionization front aligns with the minimum of the potential well (see Fig.~\ref{set-up}). 
As the electrons are born and slip backward in the co-moving frame, the electrostatic pseudo-potential is converted into kinetic energy, accelerating them. 
If the potential is sufficiently large, it will facilitate trapping of electrons in the plasma wave.

The trapping condition can be derived from the Hamiltonian of an electron initially at rest (\(v_\mathrm{i} \approx 0\)) in a blowout propagating with phase velocity \(v_\phi \approx c\)~\cite{PakPRL2010PhysRevLettshort}.
In contrast to the convention used in Ref.~\cite{PakPRL2010PhysRevLettshort}, where the pseudo-potential is maximized in the blowout center, we adopt a convention where \(\Psi\) has a minimum at the zero crossing similar to \cite{Oz2007IonisationInjection}.
Accordingly, an electron liberated at co-moving coordinate $\zeta_\mathrm{inj}$, where the pseudo-potential is $\Psi_\mathrm{inj}$, will become trapped in the wakefield if there exists a co-moving coordinate $\zeta_\mathrm{trap}$ such that the condition $\Psi_\mathrm{inj} - \Psi_\mathrm{trap} = -m_\mathrm{e}c^2/e$ is satisfied.

The range of injection coordinates that lead to trapping in the wakefield is given by the condition $\Psi_\mathrm{inj} - \Psi_\mathrm{max} \leq -m_\mathrm{e}c^2/e$, indicated by the gray-shaded region in Fig. \ref{set-up}. This relationship between injection position and \(\Psi\) independently determines the final co-moving trapping location within the plasma wave for cold released electrons, thus fixing the electric field by which the electrons are subsequently phase-constantly accelerated.

The parabolic symmetry, together with the localized injection dynamics governed by the electrostatic potential and the decoupled nature of the plasma photocathode, form key foundations of the stability mechanisms discussed throughout this paper. The primary fluctuations and their contributions to the stability process investigated in the following sections are the electron driver energy, energy spread, and current, as well as the plasma photocathode injector laser energy and its spatiotemporal synchronization with the driver.

\section{Witness charge is fully decoupled from driver beam charge}\label{SectionWitnessChargeDecoupled} 
To reflect shot-to-shot jitter in this setup, we perform simulations with significant variations in the driver beam density by varying its charge $Q_\mathrm{D}$.
In contrast, the released and trapped witness charge remains practically 
completely independent of drive beam charge fluctuations.
This insensitivity arises because the witness charge yield is determined by the plasma photocathode injector laser field strength and the helium density, which together set the number of electrons that can be released per unit volume from He$^{+}$ ions. 

With a baseline dimensionless light amplitude $a_0 = eE_{\mathrm{L}}/(\omega_{\mathrm{L}} \mathrm{m}_e c) \approx 0.0611$, where $E_{\mathrm{L}}$ is the peak amplitude of the laser's electric field and $\omega_{\mathrm{L}}$ is the laser frequency, the pulse will tunnel ionize the He$^{+}$ ions and will release He$^{2+}$ electrons with $Q_\W \approx  4.9$ pC of charge.

\begin{figure}[!htbp]
\includegraphics[width=0.45\textwidth]{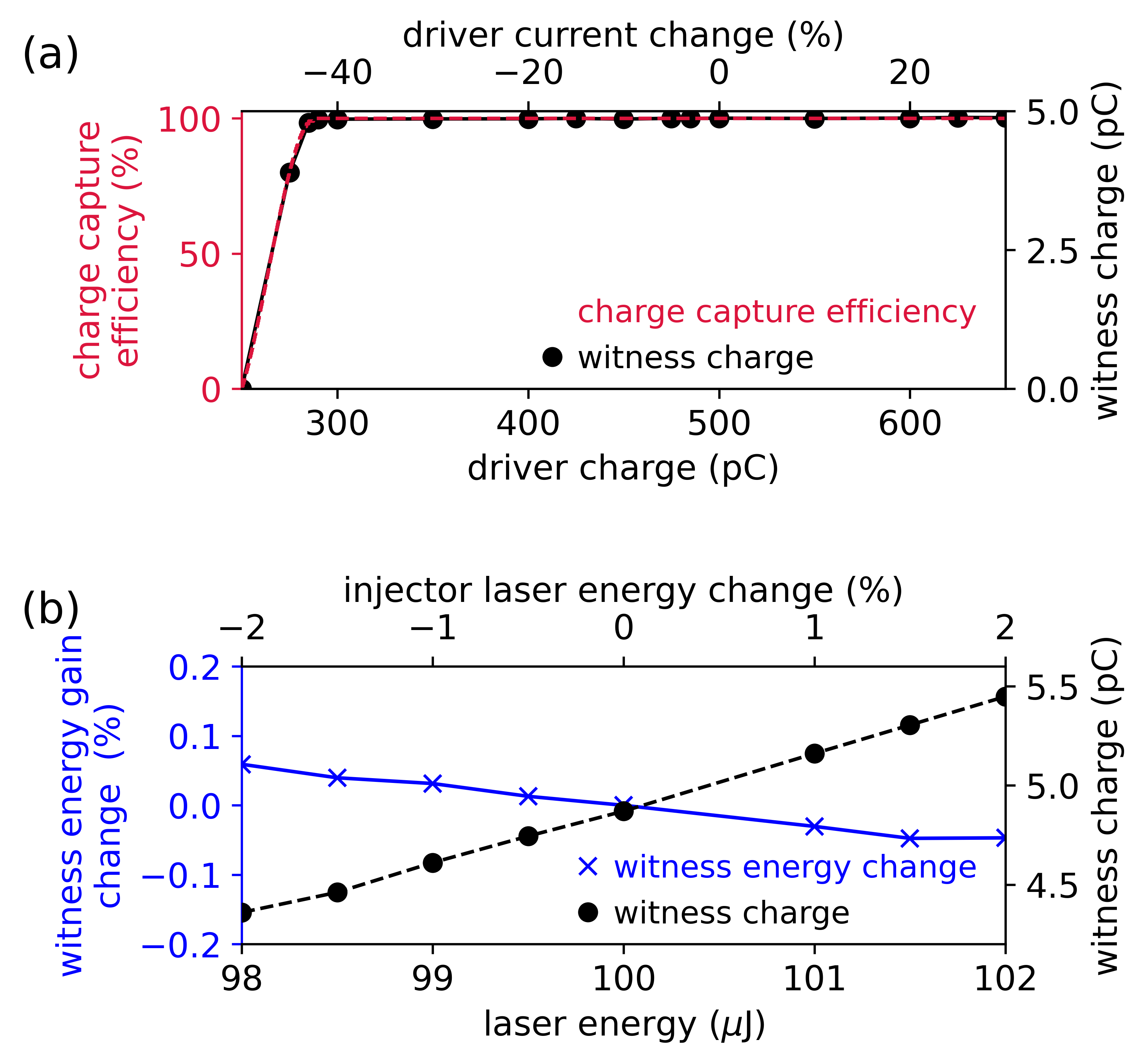}
\caption{Witness charge when driver charge or laser energy varies.
The stability of the injected witness charge versus varying driver charge and current is shown in panel (a) at constant laser energy, while panel (b) shows witness energy stability versus injector laser energy  variation at constant, baseline driver beam charge. } 
\label{fig: dr_charge-las_energy}
\end{figure}

The simulation results shown in Fig.~\ref{fig: dr_charge-las_energy}\thinspace(a) reveal that a witness charge capture efficiency of 100\% is achieved, as long as the driver charge amounts to $Q_\mathrm{D} \gtrsim  290$ pC, corresponding to a current of $\sim5.8$ kA, the threshold where the electrostatic potential becomes large enough to trap all electrons released around the blowout center.

In principle, the wakefields themselves may also contribute to the charge yield through superposition with the laser pulse field, leading to a local increase in the electric field amplitude.
However, releasing at the blowout center has the concurrent advantage that not only is the electrostatic potential minimized, 
but the wakefields are nearly zero in this region~\cite{YunfengPhysRevSTABshort}.
Therefore, the wakefield has a negligible impact on the charge yield. 
Furthermore, since this showcase uses the second ionization level of helium as the 
HIT level -- in contrast to the first ionization level of He used in, e.g., Ref.~\cite{Deng2019Trojan} -- the laser intensity is chosen to be significantly higher.
As a result, the contribution from shot-to-shot wakefield variation becomes even more marginal.
In any case, the produced witness bunch charge is fully decoupled from the driver beam charge, and respectively, current, over a very wide range.  

While charge capture efficiency and stability are natural outcomes of the laser-dominated collinear hybrid LWFA $\rightarrow$ PWFA plasma photocathode setup, these factors present significant challenges in other LWFA or PWFA injection and staging schemes.
For example, in plasma density downramp injection in LWFA \cite{GoetzfriedPRX2020} or PWFA \cite{OptimizedDownrampAlbertoPhysRevAccelBeams.20.091301,Ullmann2021Torch,Couperus2021PWFADensityDownramp,HUE-2023-PWFA}, the injection rates are strongly dependent on the strength of the driver, making them highly sensitive to fluctuations of the driver beam from shot-to-shot.  

Similarly, during LWFA staging~\cite{SteinkeNature2016} or external injection~\cite{WuStagingNature2021}, full charge capture is very challenging, arising from a high sensitivity to alignment between the trajectory of an already relativistic electron beam and the subsequent wake~\cite{Pronold2018-external}.
As a result, spatiotemporal fluctuations and/or angular offsets of the injected electron beam with respect to the wake have far-reaching consequences for the subsequent acceleration dynamics.

While driver beam electron density is the key parameter for wakefield excitation, the corresponding critical parameter for the plasma photocathode injector is the laser intensity.  
It is therefore worthwhile to investigate the witness beam charge when the laser energy is varied, but the driver beam charge density is assumed constant. 
  
When the injector laser energy is varied by $\pm 2$\% around this baseline, the corresponding tunneling ionization rates produce witness bunches with charge variation of 4.9$\pm0.5$ pC~($\pm 10$\%) as shown in Fig. \ref{fig: dr_charge-las_energy}\thinspace(b), right $y$-axis.
The slightly different charge has a very small effect on the energy gain, which due to beam-loading is smaller for higher laser energy than for lower laser energy (see Fig. \ref{fig: dr_charge-las_energy}\thinspace(b), left $y$-axis).

The plasma photocathode laser pulse can either be provided as an intrinsically synchronized pick-off from the LWFA laser system, or delivered via an external, synchronized laser arm. 
Since a laser energy variation of $\pm 2$\% is significantly worse than the operational standard achieved by commercial high-power laser systems shot-to-shot, this parameter should be regarded more as a tunable knob for adjusting  the witness beam charge -- and consequently compound quality metrics such as the witness bunch brightness \cite{Manahan2017} -- rather than a source of jitter.  
Moreover, various techniques may be explored to normalize the pulse energy at the sub-mJ level by sacrificing input pulse energy, such as cross-polarized wave normalization, reverse saturable absorbers \cite{Harter1984-reverse-saturable-absorber}, non-linear optical limiters \cite{Matsuo2012-OKE-laser-stabilization-non-linear-opical-kerr-effect} etc.

\section{Witness energy gain is independent of driver beam energy}\label{SectionWitnessIndependentOfDriverEnergy}
Next, we investigate the impact of energy fluctuations of the LWFA-generated driver beam, by varying its central energy from 125 MeV to multi-GeV levels, whilst keeping its current, transverse properties and the injection laser properties fixed. 
\begin{figure}[!htbp]
\includegraphics[width=0.45\textwidth]{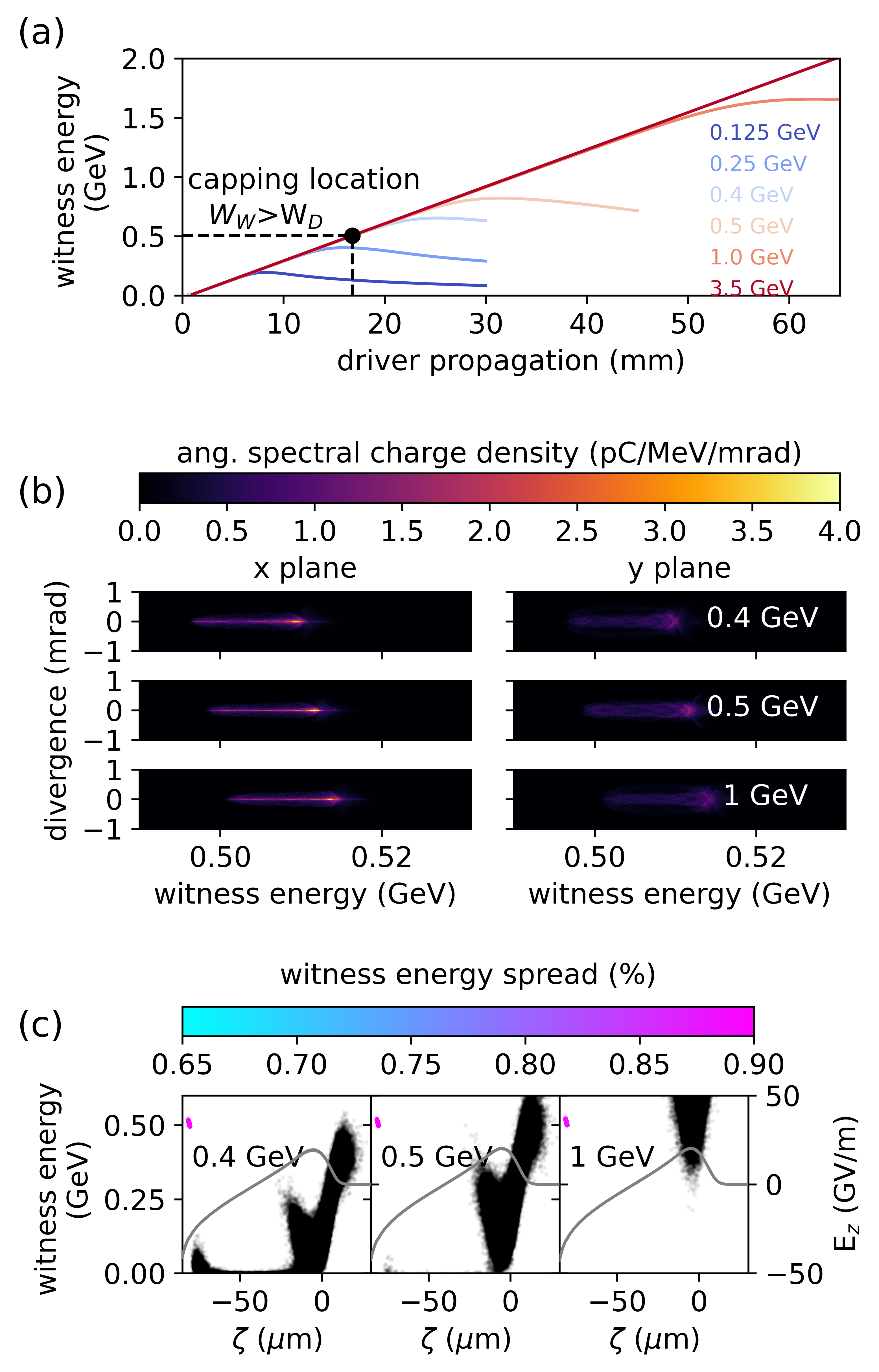}
\caption{Impact of driver beam electron energy (0.125–3.5 GeV) on witness energy gain. Panel (a) shows the energy gain of the witness beam as the driver beam energy varies. Panel (b) presents the spectral charge density projections of the witness beam in both the $x$ and $y$ planes for driver energies of 0.4 GeV, 0.5 GeV, and 1 GeV, at a potential cut-off location indicated by the dashed line in panel (a). Panel (c) displays the energy phase space plots of both the driver (black) and witness beams (left $y$-axis), color-coded by relative energy spread, corresponding to the conditions in panel (b). The longitudinal accelerating on-axis field is shown as a solid grey line (right $y$-axis).}
\label{energy-stable}
\end{figure}

Figure \ref{energy-stable}\ts(a) reveals that the witness energy gain rate is linear, and nearly constant across the driver parameter sweep,  
despite the order of magnitude variation in driver energy.
Fundamentally, this is because at sufficiently high electron energies $W_\mathrm{kin}$ and corresponding Lorentz factors $\gamma = 1 + W_\mathrm{kin}/(\mathrm{m}_e c^2)$, the individual electron velocities $v$ are all very similar and close to $c$, due to $v = c \sqrt{1 - \left(1/\gamma^2 \right)}$.  
Therefore, the driver beam current profile remains similar, and in turn the produced wakefield.   
That said, naturally, higher energy drivers are capable to drive the wake over longer distances, and hence are able to sustain the linear energy gain of the witness to higher energies. 

The linear energy gain, enabled by the phase-stable nature of the PWFA, remains largely unaffected by variations in the transverse evolution of the driver beam -- such as betatron oscillations and lensing effects -- which are directly linked to the driver's slice energy.

The energy gain is not entirely identical, as shown in Fig.~\ref{energy-stable}\ts(b), which displays the spectral charge densities of the electron witness bunches -- representing artificial energy spectra that would correspond to experimental observables.
These plots reveal that the witness energy is slightly higher for higher-energy drivers compared to the baseline.
They also exhibit differences in transverse divergence, reflecting an asymmetry in transverse emittance between the two projection planes.
This asymmetry originates from the linear polarization of the injector laser pulse, which increases the thermal emittance \cite{schroeder2014PRSTABTrojan} of the witness electrons along the polarization direction -- here, the $y$-direction.
As a result, the normalized projected emittance is $\epsilon_{n,y} = 29.0~\mathrm{nm~ rad}$ for the baseline witness, compared to $\epsilon_{n,x} = 12.3~\mathrm{nm~rad}$ in the orthogonal direction.

Additional spectral examples, including the driver beam signal, are provided in Appendix A.
These spectra are taken at the cut-off location marked by the black dashed line in Fig. \ref{energy-stable}\ts(a), representing a potential experimental extraction point designed to obtain stabilized witness energy even under extreme experimental conditions of wildly varying driver beam energies from shot-to-shot at mean witness energies exceeding the baseline driver value ($W_W>W_D$).

In this scenario, driver beam energies in the range of $\sim450-550$ MeV yield witness beam energies with a precision better than $\pm 0.1\%$, while driver energies above $\sim 360$~MeV can maintain witness energy stability within $\pm 1\%$.
These parameter scans demonstrate that the output energy stability of the witness beam can exceed the stability of the driver beam by more than an order of magnitude. 
 
Plots of the longitudinal phase space for both driver and witness populations highlight effects that depend on the assumed relative energy spread of the driver beam, which here remains at the baseline 10\%. 
The lowest-energy electrons within the driver beam that are located in its strongest decelerating region, are slowed down most rapidly. This leads to a progressive lengthening of the driver beam, which in turn reduces its peak current and weakens the wakefield strength during driver depletion.  

Eventually, as shown in Fig. \ref{energy-stable}\ts(c), driver beam electrons are fully depleted, and slip into the accelerating phase of the wake. 
This effect, already observed in simulations \cite{PhysRevSTAB.13.101303HabsCollectiveDecel,HiddingPRL2010PhysRevLett.104.195002short} and seen experimentally \cite{Chou2016, PPena2024-driver-reacceeleration}, 
is naturally most pronounced for low-energy driver beams with large energy spreads. 
Accordingly, it is most visible in Fig.  \ref{energy-stable}\ts(c) for the case of a 0.4 GeV driver.  
As shown by the colorbar (normalized across Fig.  \ref{energy-stable} and Fig. \ref{energyspread-stable}), the (projected) relative energy spread (r.m.s.)  of the witness beams is $<1.0$\%. Compared to the driver electron beam energy spread of 10\%, this constitutes a reduction of more than an order of magnitude and signifies a quality enhancement of the extracted witness beam in terms of its transverse emittance and concurrently its energy spread, enabled by the hybrid LWFA$\rightarrow$PWFA staging.

Simultaneously, the hybrid LWFA$\rightarrow$PWFA platform therefore functions as an energy and beam quality transformer \cite{Hidding2019FundamentalsLWFA-PWFA,Alberto-Hybrid}. 
The energy and energy spread effects come with a trade-off in overall energy efficiency between the input LWFA beam and the output beam generated by the plasma photocathode in the PWFA stage. 
However, as the reproducibility of the LWFA-generated driver beam improves, the efficiency can be further optimized.
This optimization is fundamentally governed by the depletion length of the driver beam with the lowest mean energy.
As a result, a capping point closer to the maximum baseline witness energy -- approximately $\sim 800~\mathrm{MeV}$ in this case -- becomes achievable.
 
\section{Witness energy gain is independent of driver beam energy spread}\label{SectionWitnessIndependentOfDriverEnergySpread}
However, it is not only the mean energy of the driver beam that can vary from shot-to-shot.
It is well-known that the energy spread of LWFA-generated electron beams can also fluctuate significantly. 

\begin{figure}[!htbp]
\includegraphics[width=0.45\textwidth]{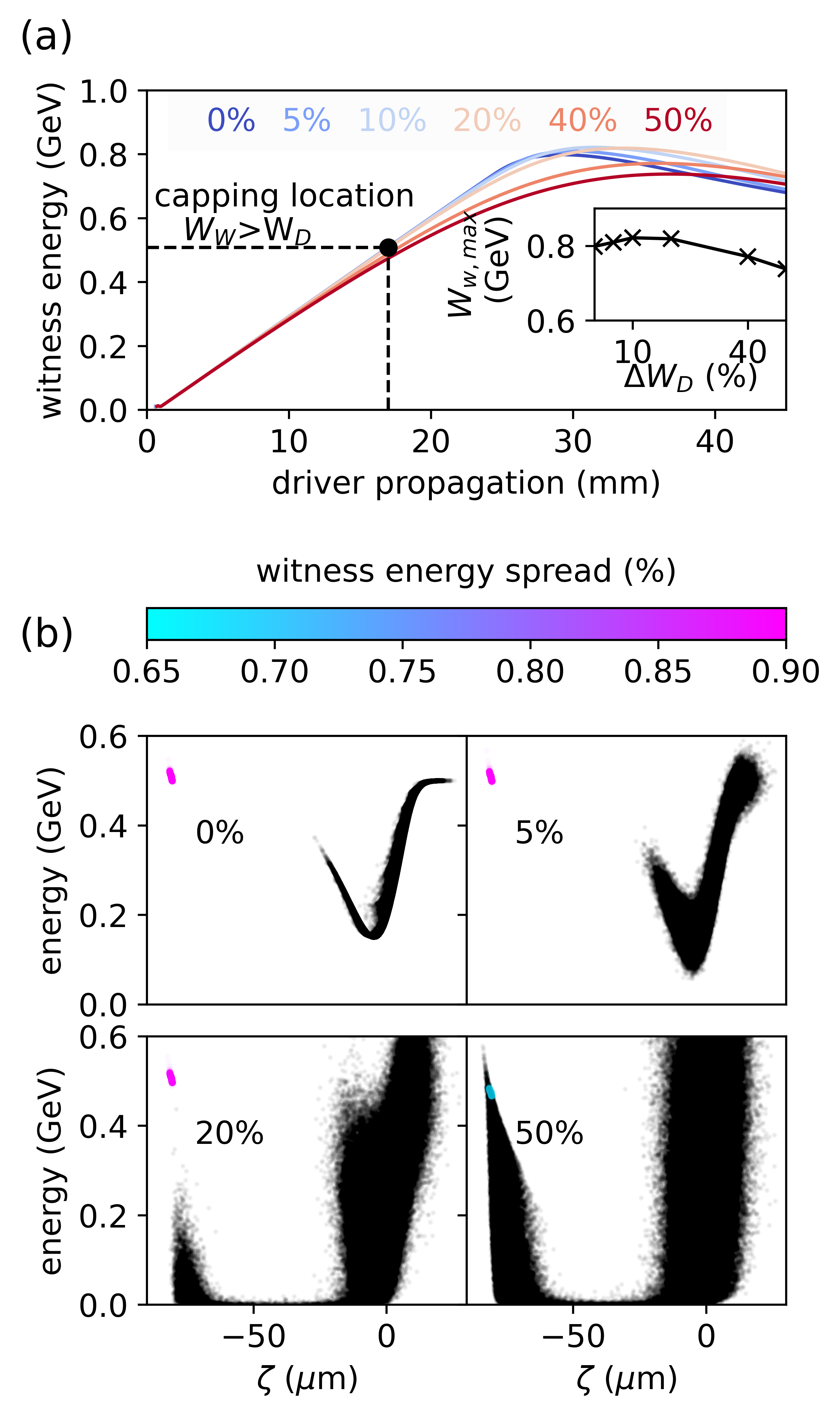}
\caption{Impact of driver beam energy spread on witness bunch energy gain. Panel (a) illustrates the witness energy gain as a function of driver energy spread for a fixed driver energy  of $W_\D$=500 MeV.  The dashed line indicates the threshold at which the witness energy exceeds the driver energy. 
The inset shows the maximum mean energy gain trend for the witness bunch. Panel
(b) presents longitudinal phase space plots for the driver and witness bunches at different driver energy spreads (0\%, 5\%, 20\%, 50\%), while the witness bunch maintains a relative projected energy spread of less than 1\%. The plots demonstrate the influence of driver energy spread on the longitudinal distribution of both bunches.}
\label{energyspread-stable}
\end{figure}

To determine the effect this may have on the plasma photocathode-generated witness bunch in the PWFA stage, a scan is undertaken by varying the driver's relative uncorrelated energy spread between 0 and 50\%,  while keeping the mean driver electron energy constant at its baseline level of $W_\mathrm{D} = 500 $ MeV. 
Fig. \ref{energyspread-stable}\ts(a) shows the corresponding witness energy gain curves with the longitudinal phase space plots of driver and witness for selected driver energy spreads displayed in Fig. \ref{energyspread-stable}\ts(b). 

As evident from Fig.~\ref{energyspread-stable}\ts(a), the energy gain remains nearly unaffected by even large variations in the driver energy spread during the early stages of acceleration.

With increasing acceleration distance and driver beam deceleration, in plasma waves driven by driver beams with larger energy spreads such as 50\%, the witness energy gains begin to reduce, for similar reasons as discussed in the context of varying driver beam mean energies. 
Notably, lower driver energy spreads allow the witness beam to reach its maximum energy earlier.
 
While the largest energy spread drivers yield the lowest peak witness energies, drivers with moderate energy spread such as 10\% can actually achieve slightly higher maximum energy gains than the hypothetical zero-spread case ~(see Fig. \ref{energyspread-stable}\ts(a) inset). 
The acceleration distances at which these maxima occur vary non-trivially, reflecting the complex interplay of wakefield dynamics and pump depletion effects.

Again, introducing a capping point -- e.g., where the witness energy exceeds the driver energy -- helps ensure stable witness energy output, even with strongly varying driver energy spreads from the LWFA stage.
In this case, the lowest energy spread cases reach the target energy before those with higher energy spread.
At the chosen extraction point, the mean witness beam energy is 507.0~MeV. Our analysis shows that if the driver energy spread remains below 26.5\% r.m.s., the extracted witness energy stays within a $\pm$1\% tolerance, demonstrating the hybrid LWFA$\rightarrow$PWFA scheme’s broad tolerance to energy spread jitter.
Additionally, all resulting witness bunches exhibit projected energy spreads below 1\%, reducing the witness energy spread by more than an order of magnitude compared to the baseline driver beam.

In this context, it is important to note that the selected robust working point (Fig. \ref{fig: dr_charge-las_energy}) ensures that witness electron trapping occurs in a region of the wakefield with comparatively lower accelerating gradients, rather than at the wake's vertex. 
While the vertex offers higher initial wakefield amplitudes -- and thus the potential for greater witness energy gains -- trapping in the chosen region provides enhanced beam stability. 
This placement reduces the witness electrons’ vulnerability to wakefield evolution and lowers the risk of slipping out of the focusing and/or accelerating phases of the wakefield. 

Another consequence of positioning the witness bunch in a more advanced phase of the wakefield is the potential for spatial overlap with decelerated electrons from the driver beam. 
Here, they may contribute to beam-loading \cite{KatsouleasBeamLoadingPlasmAcc1987}, which may be exploited for an interesting effect in particular in context of ultralow emittance electron beams.
In Ref. \cite{Manahan2017}, it has been found that beam-loading by an `escort' beam released to overlap with an ultralow emittance witness beam that has already been accelerated to sufficiently relativistic energies, can reduce or even compensate the witness energy chirp, without spoiling its emittance. 
In the present configuration, such beam-loading can be provided by the decelerated drive beam fraction that slips back into the accelerating phase of the wakefield. 
Fig. \ref{energyspread-stable}\ts(b) shows the signature of this effect in the snapshot for the scenario of the 50\% driver energy spread. Here, the witness' projected energy spread is significantly lower ($\sim0.68$\% vs. $\sim0.9$\%) than in the other three snapshots shown, as the result of beam-loading (see Appendix B). 
This creates an opportunity to purposefully reduce the witness beam's energy spread while increasing its energy, provided there is sufficient control over the driver beam. 
While this scenario requires further exploration, it could be a natural and perhaps rather unique fit to hybrid LWFA$\rightarrow$PWFA-plasma photocathodes, since they can provide sufficiently large energy spread driver beams, and ultralow emittance witness beams sitting in sufficiently advanced phases of the wakefield for spatial overlap with decelerated driver beam electrons.   
Improving the energy spread while keeping the low emittance could be crucial for applications such as free-electron-lasers, which impose strict thresholds on both parameters 
\cite{Habib_2023_NC_AttosecondAngstromFree,Galletti2024}.  

\section{Witness production has high spatio-temporal release tolerance}
We now examine the influence of the spatio-temporal release position of electrons by the plasma photocathode within the plasma wave.

\begin{figure}[!htbp]
\includegraphics[width=0.45\textwidth]{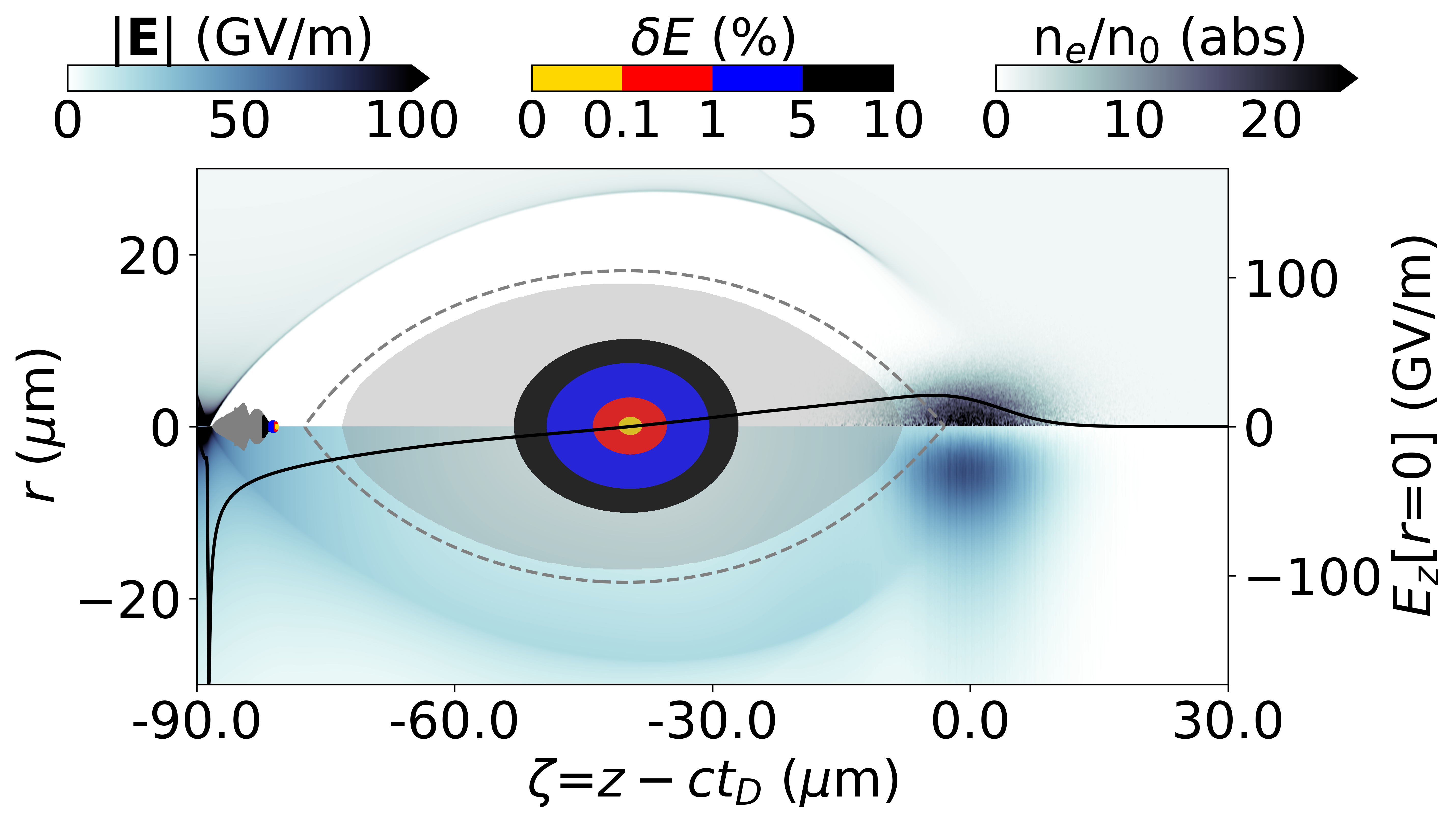}
\caption{Intrinsic robustness of witness bunch production against photocathode laser spatio-temporal jitter, shown as a function of longitudinal and radial release position within the blowout.
Plasma and driver electron densities are shown in a white-light blue-green gradient, and the localized electric field magnitude is depicted in a white-dark blue gradient. Electrons released within the color-coded energy bands around the 2D trapping ellipse (grey area) are trapped in small volumes at the back of the blowout. The dashed grey line represents the 1D theoretical trapping ellipse, with energy deviations (\(\delta E\)) marked by contour boundaries and the on-axis longitudinal electric field shown as a solid black line.
}
\label{spat-temp}
\end{figure}

As before, the baseline driver beam is assumed. 
A test-particle approach is employed to simulate all possible electron release positions simultaneously, independent of specific laser parameters. 
The results show that electrons released within the solid grey ellipse (see Fig. \ref{spat-temp}) are captured with 100\% capture efficiency, accelerated, and retained throughout the entire acceleration length. 

As expected, this aligns with the analytical solution (dashed grey) for trapping from negligible initial velocity, corresponding to the region satisfying the simplified trapping condition discussed earlier, here applied in 2D.

Application-specific witness bunch energy stability targets then inform the acceptable spatiotemporal plasma-photocathode release region in the wake as a subset of the trapping ellipse.
Iso-contour lines show where released, and subsequently trapped and accelerated electrons would obtain $0.1\%$ (yellow), $1.0\%$ (red), $5.0\%$ (blue), $10.0\%$ (black) energy deviation with respect to electron release at electrostatic potential minimum. 
The symmetry of the injection acceptance energy iso-contours reflects the underlying symmetry of the electrostatic potential around the center of the wakefield.
In hybrid LWFA$\rightarrow$PWFA configurations, the temporal synchronization jitter of the injection laser may be less critical than its pointing jitter. To achieve the 1\% witness energy stability target considered in this study, the transverse shot-to-shot pointing fluctuation must be limited to below $\pm 3\,\mathrm{\mu m}$—a level of stability within reach of emerging high-power laser system capabilities~\cite{MaierPRX,WU2020-TW-pointing-stability,Amodio_2025_pointing_ML}.

The positive gradient of $E_z$, in concert with linear focusing fields, $E_r$, facilitates automatic longitudinal and transverse compression of electrons at the end of the plasma wave bucket.
The real space volumes of trapped test electrons corresponding to the release volumes are indicated with the same  colors as the target iso-contour lines in Fig. \ref{spat-temp}. 
The auto-compression is a particularly beneficial feature of the aforementioned injection and trapping process, automatically generating witness bunches with ultrashort bunch durations and multi-kA peak currents, exceeding 2.2 kA in the showcase configuration. 

When injection occurs off-axis, betatron oscillations are induced by the focusing fields; however, these oscillations are rapidly damped upon acceleration given the increase in particle $\gamma$ with time. Effectively, the witness bunch follows the orbit of the driver beam. This offers a significant advantage over external injection schemes involving already relativistic witness beams, where even minor transverse offsets or pointing errors can severely compromise the acceleration process \cite{Pronold2018-external}. 
Recent studies have indicated \cite{Maity2025-external-injection-lwfa-offsets} that already micrometer-scale emittance growth becomes difficult to control with even sub-micrometer misalignment. 
In contrast, simulation studies suggest that plasma photocathodes can limit emittance growth to the nanometer level -- more than two orders of magnitude lower -- even in the presence of substantially larger transverse offsets \cite{Habibplasma-photocathodes-2023}.

\section{Witness beam energy gain self-adapts to driver charge density fluctuations}

In section \ref{SectionWitnessChargeDecoupled}, the driver charge density was varied to demonstrate the witness beam injection rates in the center of the wake are independent of the driver charge density, and the witness charge yield is hence fully decoupled from shot-to-shot-variations of the driver beam. 
However, as is well-known from theory \cite{lotovPRE2004blowout,Lu2006,GolovanovPhysRevLett2023}, experiments and even directly observable via shadowgraphy \cite{SchoebelNJP2022}, variations in the driver density relative to the plasma density $n_\mathrm{D}/n_\mathrm{p}$ due to driver charge $Q_\mathrm{D}$, length $\sigma_\mathrm{z}$, transverse size $\sigma_\mathrm{r}$  or shape variations from shot-to-shot, can also strongly vary the peak wakefield amplitude, the plasma wavelength and hence the plasma wave cavity size. 
Therefore, while the accelerating wakefield is only marginally dependent on driver beam  energy and its energy spread as discussed in sections \ref{SectionWitnessIndependentOfDriverEnergy} and \ref{SectionWitnessIndependentOfDriverEnergySpread}, in contrast it is strongly dependent on driver charge density variations \cite{Moritz2022-PRX}.

It is also well-known that injection rates in LWFA are difficult to control \cite{Cho2018, ZhangPRABfacet-iiDownramp}, and the slice and total current values of electron beams produced by LWFA therefore often exhibit even larger shot-to-shot variations than in energy or energy spread.
While total charge is easy to measure, variations in electron beam length and shape are notoriously difficult to quantify. 
Consequently, shot-to-shot variations in LWFA-produced driver beam charge density represent a very significant, yet often elusive and challenging-to-measure, source of variation for wakefields in hybrid LWFA$\rightarrow$PWFA, ultimately influencing the energy gain experienced by witness beams injected in the PWFA stage.
For example, even assuming a perfectly synchronized, externally injected, pre-accelerated  relativistic electron witness beam, its energy gain in the PWFA stage would vary significantly from shot-to-shot. 
Such a perfectly synchronized witness beam could be placed at a fixed distance behind the driver beam, but nevertheless would sample significantly different wakefield amplitudes and energy gains due to the variation in driver strength from shot-to-shot. 
Moreover, this scenario is hypothetical, because the typical temporal jitter in the few experiments that have been injecting electron witness beams from a linac into laser driven plasma wakefields  \cite{Wu2021-external-injection}, or in turn using lasers for injecting into plasma wakefields driven by an electron beam from a linac, is of the order of $\pm100$ fs \cite{Deng2019Trojan}. 

\begin{figure*}[!htbp]
\includegraphics[width=0.9\textwidth]{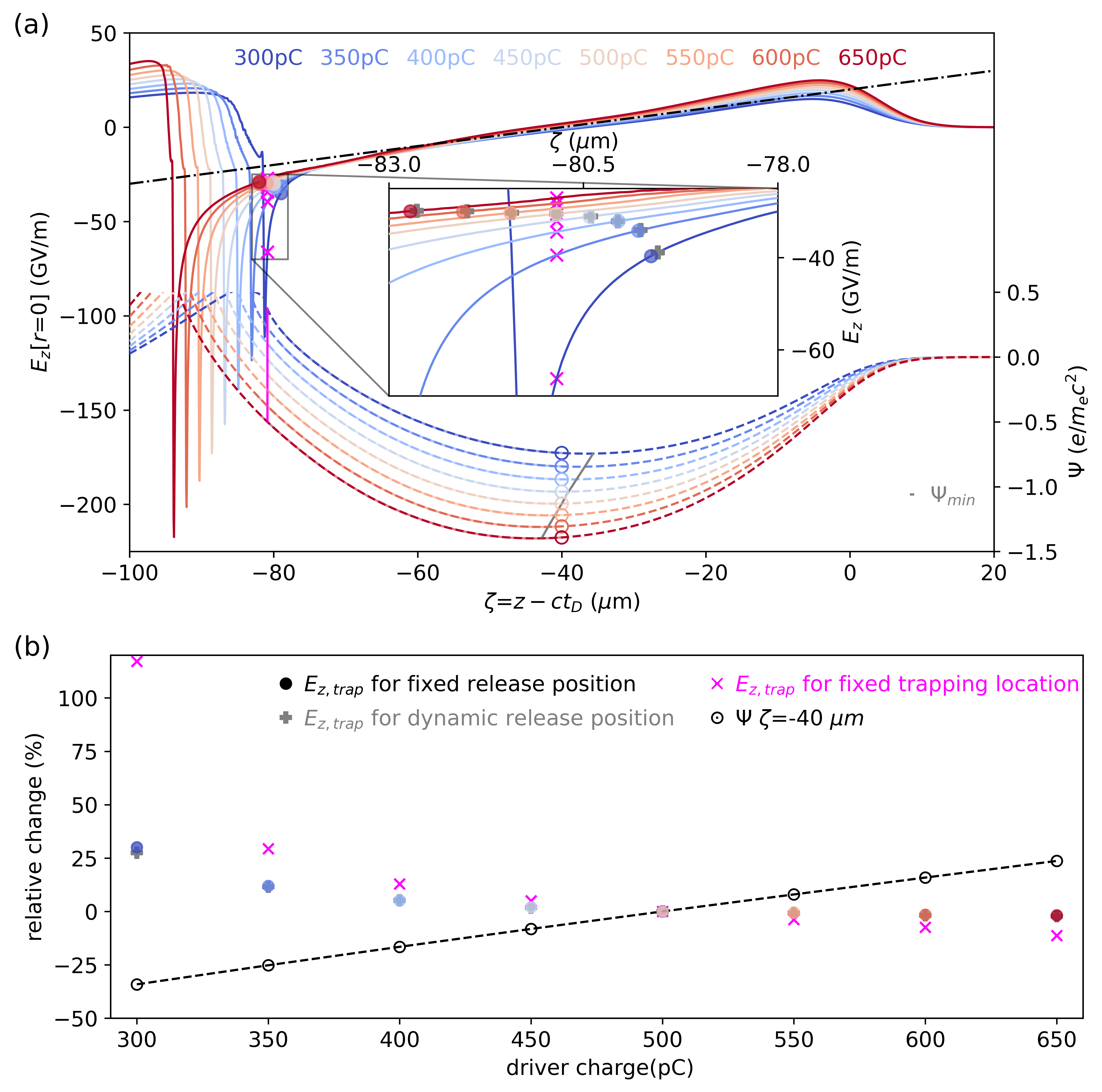}
\caption{Investigation of self-adaptive stabilization mechanisms arising from driver density-induced wakefield jitter.
Panel (a) presents wakefield line-outs for various driver charges, highlighting changes in the electrostatic potential and shifts in the location and magnitude of the minima for both $E_z$ and $\Psi_{\mathrm{min}}$.
The dash-dot black line denotes a theoretical normalized slope of 1/2, toward which very high-current drivers converge.
Overlaid are color-coded hollow circles indicating fixed injector release positions, with corresponding trapping locations marked by semi-transparent solid circles of the same color.
Grey crosses represent injection at local $\Psi_{\mathrm{min}}$ positions, while magenta crosses correspond to injection at fixed trapping locations.
Panel (b) compares the relative changes in witness behavior: releasing electrons at a fixed location near the blowout center closely matches injection at the local $\Psi_{\mathrm{min}}$, whereas $E_z$ is significantly larger when the trapping location is fixed.
}
\label{fig: wakefield physics}
\end{figure*}

In contrast,  in the hybrid LWFA$\rightarrow$PWFA scenario, we can realistically assume charge release occurs at a fixed distance behind the driver beam, due to the intrinsic synchronization with the injector laser pulse. How does the witness energy gain change in this scenario as consequence of driver beam density jitter? 
To investigate this, we vary the drive beam density as in section \ref{SectionWitnessChargeDecoupled}, and set the plasma photocathode release position to the electrostatic minimum for the case of the driver charge of $Q_\mathrm{D}=500$~pC. 
The variation of charge from $Q_\mathrm{D} =300$~pC to 650~pC 
corresponds to a relative change of -40\% to +30\%.

Fig.  \ref{fig: wakefield physics}\ts(a) visualizes the significant variation in electrostatic potentials and longitudinal wakefield configurations as a response to changes in driver charge (density). 
The color-coding emphasizes that wakefields are stronger (red) and weaker (blue) for driver charge higher/lower than the baseline driver charge $Q_\mathrm{D} =500$~pC. 
The color-coded hollow circles represent the plasma photocathode release position $\zeta_\mathrm{rls}$, and highlight that on the one hand the electron release positions are located at electrostatic potentials that are significantly different for the range of $Q_\mathrm{D}$ considered. 
On the other hand, for all cases other than the baseline driver charge, they are also located not at the electrostatic potential minimum of the waves, but to the right (red, stronger waves) and left (blue, weaker waves) of their individual minimum, represented by the dashed grey dotted line.
However, the positions in the wakefield  where the electrons are trapped ($\zeta_\mathrm{trap}$, semi-transparent solid circles) are closer to the release position for weaker waves (blue) than for stronger waves (red). 
This may be counter-intuitive, because in a constant wakefield, release positions outside the electrostatic potential minimum result in trapping positions $\zeta_\mathrm{trap}$ farther behind in the plasma wave. 
In phase-constant PWFA, the wakefield at the trapping position $\zeta_\mathrm{trap}$ then determines the energy gain.

These findings highlight an intrinsic stabilization mechanism inherent to the plasma photocathode injection scheme, rendering it particularly advantageous in the context of charge variability characteristic of hybrid LWFA$\rightarrow$PWFA systems.
 
The inset in Fig.  \ref{fig: wakefield physics}\ts(a) zooms in on the trapping region, highlighting  that compared to a fixed trapping position, the range of electric wakefields at the individual trapping positions $\zeta_\mathrm{trap}$ is significantly reduced across the range of driver beam strengths. 
This represents a self-adaptive 
mechanism that stabilizes the energy gain of the witness beam despite significant fluctuations in drive beam and wakefield amplitudes.

In other words, while weaker waves have smaller electrostatic potential and peak wakefield amplitudes than stronger waves, at fixed wake phase positions, 
the  accelerating field-- and hence the acceleration rate -- is significantly \textit{larger} in weaker waves than in stronger ones. 
This is a challenge for the output energy stability of externally pre-accelerated and then injected electron beams, in addition to timing jitter.
While in the plasma photocathode mechanism, the resulting accelerating wakefields $E_{z,\mathrm{trap}}$ at individual trapping positions $\zeta_\mathrm{trap}$ are also larger for weaker waves than for stronger ones, the difference is significantly mitigated:  
trapping positions $\zeta_{\mathrm{trap}}$ automatically shift to later positions within the wake for stronger waves -- where $E_{z,\mathrm{trap}}$ is larger than at fixed trapping position. 
This dynamic shift partially compensates for variations in driver beam strengths from shot to shot, thus providing intrinsic stabilization of the  actual witness acceleration gradient, and hence the witness energy gain.
In further detail, the inset in Fig.  \ref{fig: wakefield physics}\ts(a) shows that as the driver charge is linearly increased, the variation of  $\zeta_\mathrm{trap}$ increases, while the change in $E_{z,\mathrm{trap}}$ decreases non-linearly. 
This leads to the bow-shaped distribution of wakefield vs. trapping position displayed in the inset.

Fig.  \ref{fig: wakefield physics}\ts(b) quantifies the witness energy stabilization by comparing the accelerating electric field at the resulting trapping positions $E_{z,\mathrm{trap}}$ of beams released by the plasma photocathode with those that would be obtained at a fixed  position in the wakefields. 
The $y$-axis shows the relative change compared to the baseline driver case of $Q_\mathrm{D} =500$ pC. 
The magenta plot represents the fixed wakefield
position case, i.e. the hypothetical scenario of external injection with perfect synchronization. 

The colored circles show the electric fields at trapping positions resulting from plasma photocathode emission at a fixed release position. 
The dashed grey line shows the slightly different electric fields at trapping positions when release occurs at the individual potential minima of each driver-generated plasma wave.

The hollow black circle connected by dashed black line shows the change of electrostatic potential minimum. 
This comparison highlights the intrinsic energy stability advantage of electron beams that can be expected from plasma photocathode injectors. 

The \textit{self-stabilizing} behavior of the witness energy -- and its robustness to increasing driver current, which leads to saturation of the energy gain in the high-current limit -- can be understood from first principles. 
Consider an accelerating electric field with constant slope, \(g_z \equiv dE_z/d\zeta > 0\), defined along the co-moving coordinate \(\zeta = z - ct_D\). 
We assume that the electron is released at the zero crossing of the parent wakefield, \(\zeta_0\), corresponding to the center of the blowout region where \(E_z(\zeta_0) = 0\). 
The field is then given by
\begin{equation}
E_z(\zeta) = g_z \,(\zeta - \zeta_0).
\end{equation}

Applying the trapping condition, \(\Psi(\zeta_0) - \Psi(\zeta_f) = - mc^2 / e\), and using \(E_z = \partial_\zeta \Psi\), integration yields the trapping offset
\begin{equation}
\zeta_f - \zeta_0 = -\sqrt{\frac{2 m_\mathrm{e}c^2}{eg_z}}.
\end{equation}
The accelerating field at the trapping location then evaluates to
\begin{equation}
E_{z, \trap} = -E_0 \sqrt{2\hat{g_z}}, \label{eq:ez}
\end{equation}
where \(E_0 \equiv m_\mathrm{e} c^2 k_\mathrm{p} / e\) is the cold wavebreaking field, and \(\hat{g_z} \equiv g_z  / (E_0 k_p)\) is the normalized slope of the wakefield. This result shows that the final accelerating gradient depends solely on \(\hat{g_z}\).

As established by theory and simulations \cite{Lotov2004,Lu2006PRL}, the parameter $\hat{g_z}$ asymptotically approaches 1/2 for high-current drivers ($\gg 10$ kA). In the regime of moderate currents (around $\sim 10$ kA), $\hat{g_z}$ exhibits only a weak dependence on the driver's current, typically settling near 0.3, in agreement with earlier findings \cite{albertodelaOssa2015}. 

Due to the square-root dependence in Eq.~\ref{eq:ez}, the influence of current on $E_{z, \mathrm{trap}}$ is further diminished, resulting in a slow convergence toward $E_{z, \mathrm{trap}} \rightarrow -E_0$ as the beam current increases. This behavior is illustrated in Fig.~\ref{fig: wakefield physics}\ts(a), where the dash-dot black line indicates the theoretical reference slope of 1/2. As shown, both $\hat{g_z}$ and $E_{z, \mathrm{trap}}$ tend to flatten and converge toward constant values for higher-charge drivers.

For lower driver currents, trapping occurs increasingly deeper within the non-linear blowout region, where deviations from linear theory become significant. 
Nevertheless, the injected beam remains more stable compared to trapping at a fixed location, further illustrating the resilience of this injection mechanism.

\section{Discussion and outlook}
The plasma photocathode is a path towards electron beams with ultralow emittance $\epsilon_\mathrm{n}$ of the order of tens of nm rad in both planes, kA-level current $I$ and in turn ultrahigh electron beam brightness $B_\mathrm{5,n}=2I/(\epsilon_\mathrm{x,n}\epsilon_\mathrm{y,n})$. 
While not discussed above, also in the present study the witness beam brightness exceeds the driver beam brightness by orders of magnitude, and hence acts as a brightness transformer for PWFA. 
The projected 5D-brightness consistently reaches values of  $B_\mathrm{5,n}>10^{19}\mathrm{A\,m^{-2}\,rad^{-2}}$, and projected 6D-brightness of the order of $B_\mathrm{6,n}> 10^{18}\mathrm{A \,m^{-2}\,rad^{-2}}/0.1\%\mathrm{BW}$. 
Although emittance, brightness and energy spread are fundamentally decisive parameters for applications such as light sources, the stability and reproducibility of electron beams are also important. Plasma photocathodes can be realized with driver beams from conventional linacs \cite{Deng2019Trojan},
but as of today very few linacs offer PWFA capabilities, in particular high-field PWFA capabilities with plasma wakefields $> 10$ GV/m. 
Only one linac engaged in PWFA research -- SLAC FACET \cite{Deng2019Trojan, YakimenkoFacet2} -- so far offered a capability crucial for hosting plasma photocathodes, namely producing a driver beam current $\gtrsim6$ kA. 

Fortunately, LWFAs 
can  produce beams with very large currents, allowing hybrid LWFA$\rightarrow$PWFA systems to even operate in the high-field PWFA regime, and to mature from concept \cite{HiddingPRL2010PhysRevLett.104.195002short} to experimental reality in recent years  \cite{Gilljohann2019Hybrid,KurzNatComm2021,Couperus2021PWFADensityDownramp,SchoebelNJP2022,Moritz2022-PRX}.
These systems were quickly envisioned as platforms capable of hosting plasma photocathodes, with their intrinsic synchronization capability recognized as an important additional advantage \cite{Hidding2012}.
Very recently, the first plasma photocathode in a  hybrid LWFA$\rightarrow$PWFA system was realized, at HZDR's DRACO facility, utilizing a 90 degree geometry \cite{UferNutter2025-to-be-published}.

With these advances, and the global number of high-power laser installations now reaching into the triple digits, the availability of LWFA-capable facilities that can be leveraged for PWFA and plasma photocathode research has grown substantially. 
This 
makes it both timely and necessary to investigate the specific conditions and parameters governing this hybrid plasma photocathode scenario in greater depth.  

In this study, we have identified and explored principal sources of input jitter anticipated when implementing an  LWFA$\rightarrow$PWFA equipped with a collinear plasma photocathode setup.
The dominant shot-to-shot variations of plasma waves driven by electron beams from today's LWFA systems originate from fluctuations in the electron beam's energy, energy spread, and charge or current. 
Then, plasma photocathode injection jitters in these plasma waves may result from spatio-temporal jitter between the plasma photocathode laser and the electron beam driven plasma wave, and from shot-to-shot changes in the laser intensity.  

Our findings show that plasma photocathode electron injection rates are purely dependent on the injection laser, and the witness electron beam charge therefore is decoupled from the driver beam strength shot-to-shot. 
This is in stark contrast to injection methods such as downramp injection or drive beam- or wakefield-induced ionization based methods, where injection rates are strongly dependent on driver and wakefield strength. 
 
Regarding the acceleration gradient, fluctuations in the driver beam energy and energy spread exhibit negligible impact until driver depletion occurs.
Lower-energy driver beams deplete earlier; thus, limiting the length of the PWFA stage can stabilize the output witness energy and operational reproducibility, as alternative to  higher witness beam energies, but reduced reproducibility. 

Interestingly, driver beam depletion especially in case of large driver energy spreads can induce beam-loading in the witness beam trapping region, which can lead to a significant reduction of witness energy spread.

Targeting the center of the blowout for plasma photocathode release benefits from the symmetry and flatness of the electrostatic wake potential in this region. 
This symmetry provides intrinsic robustness against shot-to-shot spatio-temporal relative jitter, as small variations in the electrostatic potential at the release position result in only minimal shifts in the trapping position. 
Consequently, the associated electric fields and accelerating gradients at the trapping position remain similar shot-by-shot.
 
It shall be emphasized that an injection location in the center of the wave  is a feature uniquely available for plasma photocathode injection. 
In contrast, other internal injection methods  typically either inject in earlier (e.g. driver pulse-induced ionization injection) or in later (wakefield-induced or density downramp injection) wake phases, resulting in trapping positions further towards the rear of the wake.
In contrast, plasma photocathodes naturally enable earlier trapping positions within the plasma wave, offering significant advantages on multiple fronts.
First, the wakefields at the release position are minimal, ensuring that their contribution to the witness charge yield due to superposition with the plasma photocathode laser fields is negligible. 
Second, witness beams are trapped and accelerated in a `safe zone' sufficiently well separated from the vertex of the wake. 
On the one hand, this allows robust, full capture of the released charge, as well as avoiding witness beam vulnerability to longitudinal and transverse wake evolution and potentially even charge loss. 
On the other hand, the wakefield amplitude differences at neighbouring wake positions rather deep into the blowout are rather small, minimizing the spread of accelerating gradients in case of jitters.
Third, our study reveals that in combination with the intrinsic synchronization of hybrid systems and the minimized temporal jitter between plasma photocathode and driver beam, an automatic stabilization mechanism against driver charge variation can be harnessed. 
At lower driver beam charges the electrostatic wake potential is smaller than at higher driver beam charges; however, the concurrent shift in trapping positions significantly alleviates the spread of accelerating gradients of the witness beams -- an effect  that is favourable for witness beam energy stability.

In summary, our findings show that counter-intuitively, the high beam quality feature of plasma-photocathodes does not come at the price of reduced stability, but is, according to our analysis, concomitant with improved stability.
It is revealed that the plasma photocathode can act not only as a brightness and electron energy transformer, but also as a stability transformer. 
These benefits are made as a trade-off with 
beam charge. 
The scenario of intrinsic synchronization but significantly varying driver beam from shot-to-shot, the reality of today's LWFA systems,  is complementary to the scenario of limited synchronization but state-of-the-art driver beam stability in linac-driven PWFA systems. 

The plasma photocathode thus emerges as an ideal injection mechanism for hybrid LWFA$\rightarrow$PWFA systems, offering a unique combination of three key performance enhancements -- brightness, energy, and stability -- delivered simultaneously. 
The concurrent optimization of these beam qualities is essential for demanding applications such as driving free-electron lasers, thus making this avenue an attractive route towards compact, yet high-performance accelerator systems.

\section{acknowledgements}
B.H., D.C., A.D., T.H., A.F.H., and A.S. were supported by the European Research Council (ERC) under the European Union’s Horizon 2020 research and innovation programme NeXource, ERC Grant Agreement No.~865877. This work was supported by the Science and Technology Facilities Council (STFC) [grant number ST/S006214/1, PWFA-FEL]. A.P., T.W., and L.R. were supported by the German BMBF project 05P24PF2. This research used resources of the National Energy Research Scientific Computing Center (NERSC), a DOE Office of Science User Facility supported by the Office of Science of the U.S. Department of Energy under Contract No.~DE-AC02-05CH11231, using NERSC award HEP-ERCAP0031797. Computational resources at JUWELS~\cite{JUWELS-citation-cluster-2012} were also utilized.

D.C. performed the simulations and the analysis. D.C., T.H. and B.H. wrote the manuscript, aided by input from all coauthors. The external laser in FBPIC was implemented by T.W. B.H. supervised the project.

\appendix
\section*{Appendix A: Demonstration of Witness Stabilization Using Artificially Angularly Resolved Energy Spectra }\label{Appendix: specta}

The primary instrument currently used in all LWFA and PWFA facilities for electron bunch data collection and characterization is the dipole electron spectrometer. In LWFA, the inherent absence of any driver charge allows for a clean witness-only signal. However, in PWFA, this cannot be guaranteed, as driver electrons may cause an energy phase space overlap  with the witness signal (e.g. Fig. \ref{energy-stable}\ts(c) and Fig.  \ref{energyspread-stable}\ts(b)). 
\begin{figure}[!htbp]
\includegraphics[width=0.45\textwidth]{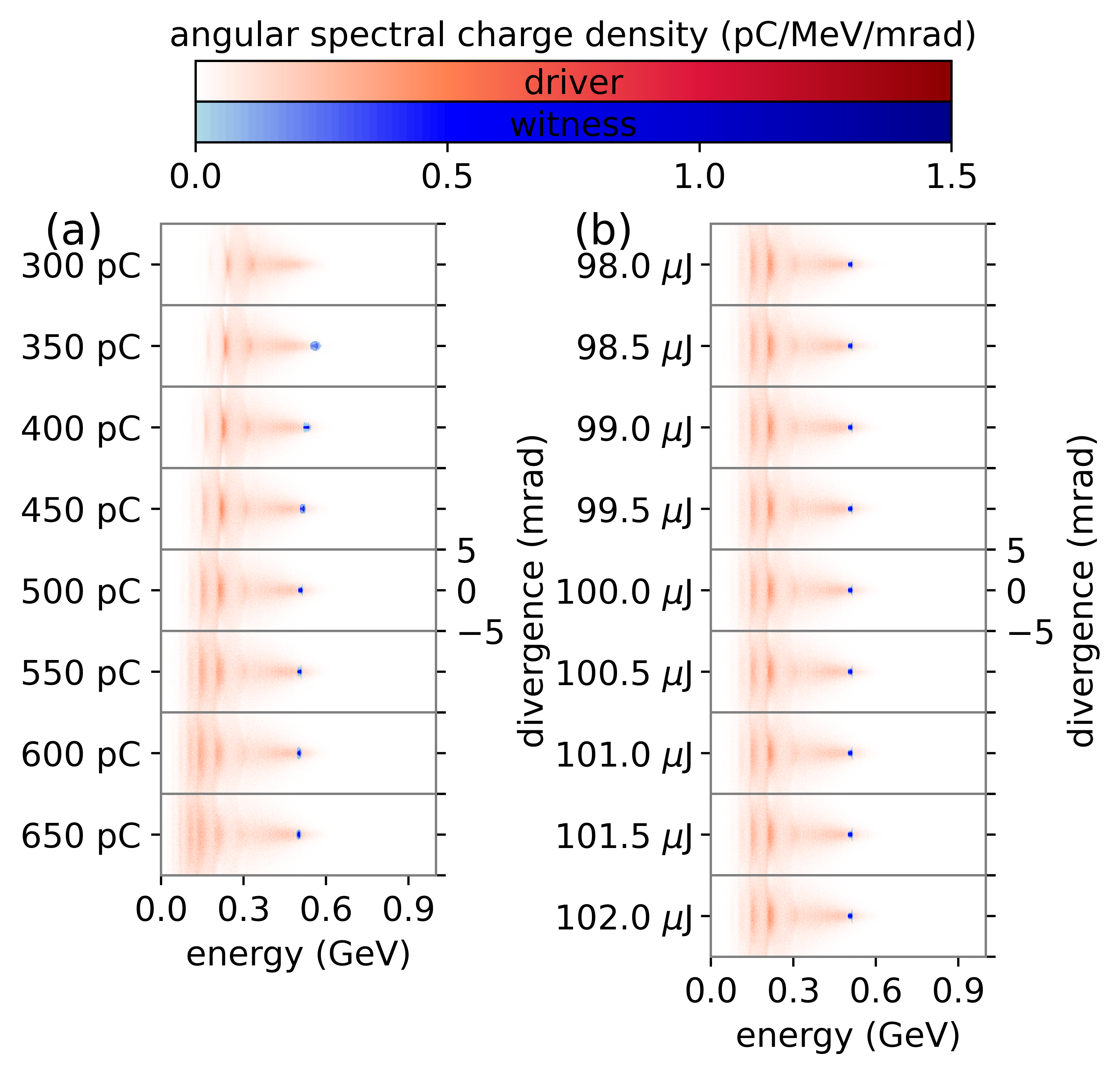}
\caption{
Exemplar artificial charge density spectra of electron beams from simulation data. Impact of driver charge variation in panel (a) and plasma photocathode laser energy variation in panel (b) on artificial charge density spectra for driver (white-red colorbar) and witness (white-blue colorbar). The driver energy and energy spread are set to 500 MeV and 10\% r.m.s., respectively. In panel (b), the driver charge is set to 500 pC. }
\label{fig: charge-laser-spectra-appendix}
\end{figure}
While the driver typically has much more charge than the witness beam, driver energy depletion and emittance and divergence growth nevertheless allows identification of the witness signal 
in almost all cases, even without background subtraction that is usually performed in experiment post-processing.

\begin{figure}[!htbp]
\includegraphics[width=0.45\textwidth]{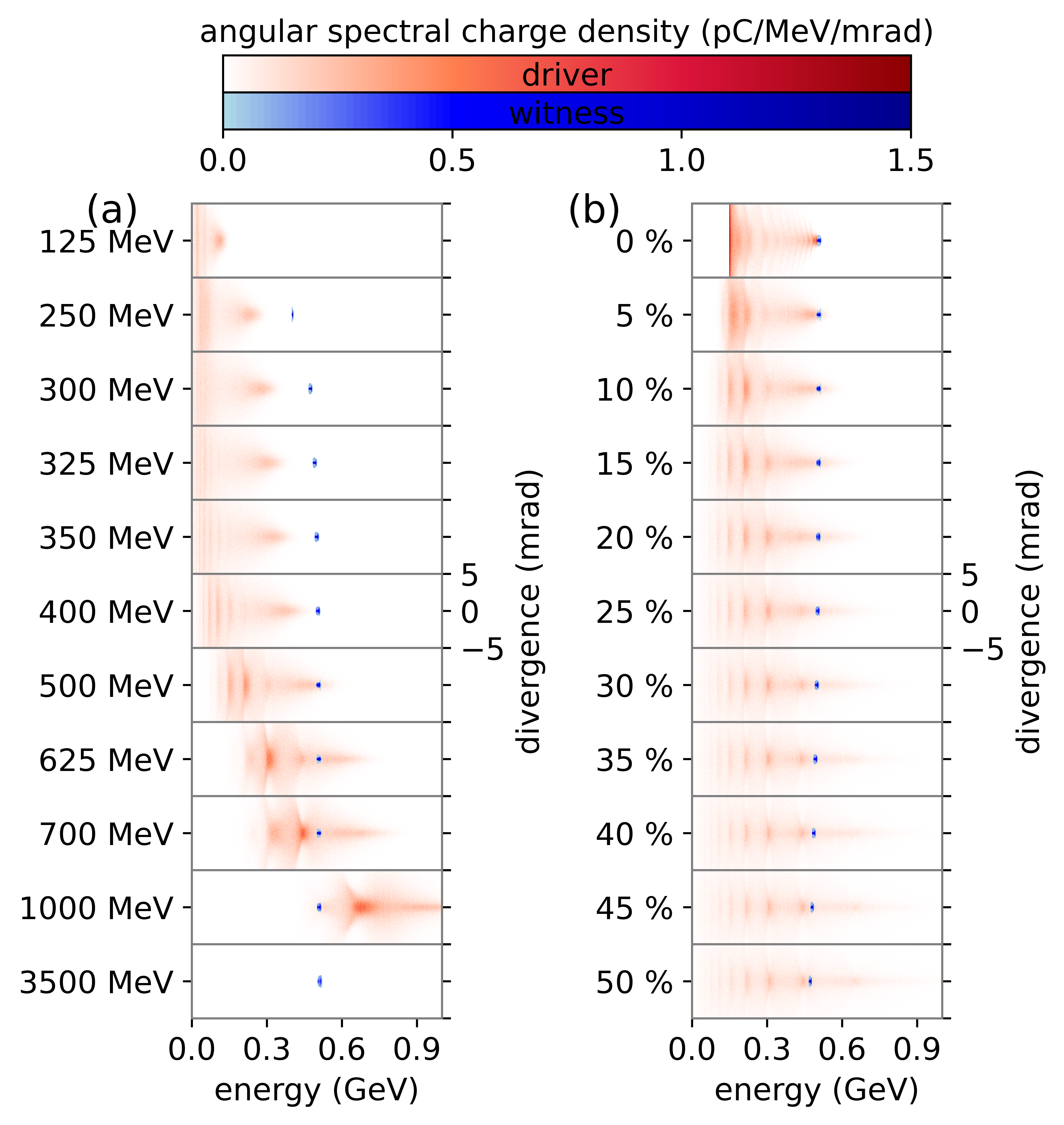}
\caption{Further artificial electron charge density spectra from simulation data. Impact of driver mean energy variation in panel (a) and driver energy spread variation in panel (b)  on artificial charge density spectra for driver (white-red colorbar) and witness (white-blue colorbar). The driver charge is set to 500 pC. In panel (a), the driver energy spread is 10\% r.m.s. In panel (b), the initial driver energy is set to 500 MeV.}
\label{fig: driver energy--appendix}
\end{figure}

Examples of the stabilization outcome are demonstrated in the form of ``artificial" spectra generated at the ``capping location", as discussed throughout the main text body of this article, shown in  Fig. \ref{fig: charge-laser-spectra-appendix} 
and Fig. \ref{fig: driver energy--appendix}. 

Fig. \ref{fig: charge-laser-spectra-appendix}\ts(a) shows the case of variation of the driver charge as discussed in context of Fig. \ref{fig: dr_charge-las_energy}\ts(a) and Fig. \ref{fig: wakefield physics}. 
The spectra show the witness energy is stable, despite the large driver charge variation. 

Fig. \ref{fig: charge-laser-spectra-appendix}\ts(b) shows artificial spectra corresponding to the case of variation in laser energy as discussed in Fig. \ref{fig: dr_charge-las_energy}\ts(b).  All other parameters kept the same, the  minimal alteration in injected witness charge produces almost identical  signal on the artificial spectrometer.

The case of energy variation of the driver beam is shown in Fig. \ref{fig: driver energy--appendix}\ts(a), amending the discussion in context of Fig. \ref{energy-stable} in the main text.  
This visualization highlights the energy stability of the witness beam, even as the driver beam transitions through regimes where its electrons are either all less energetic than the witness electrons, partially overlap in energy due to deceleration, or are entirely more energetic than the witness beam.

Fig. \ref{fig: driver energy--appendix}\ts(b) illustrates the effect of variations in the driver energy spread, complementing the discussion of Fig. \ref{energyspread-stable} in the main text and emphasizing the robustness of the witness beam against such variations.
All figures substantiate the fact that the witness beam divergence is much lower than the driver beam divergence, and that its angular spectral charge density -- an experimentally robust observable -- is enhanced. 

\section*{Appendix B: Beam loading and witness chirp reduction by driver beam depletion}\label{Appendix-beamloading}

\begin{figure}[!htbp]
\includegraphics[width=0.45\textwidth]{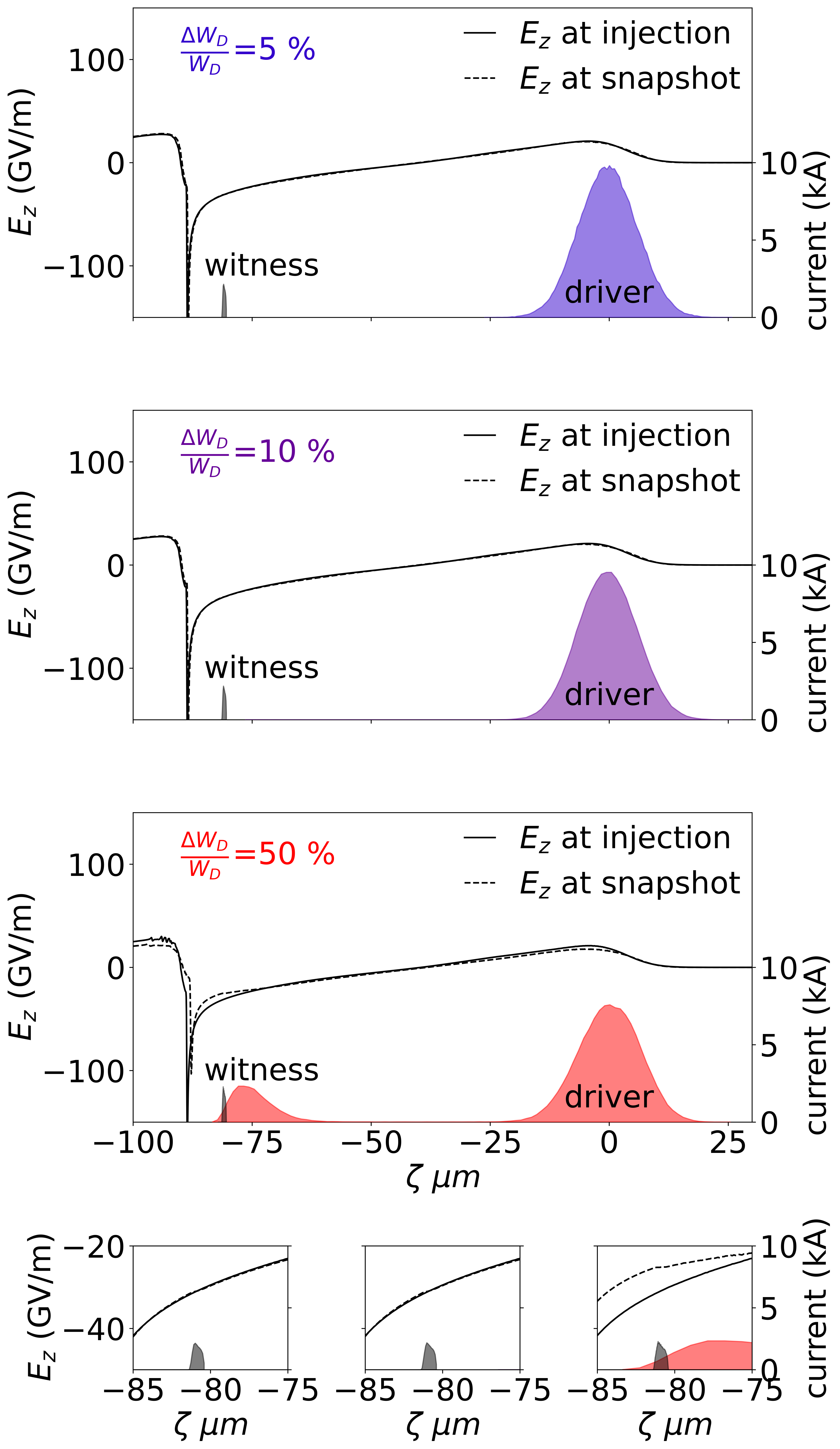}
\caption{Wakefield evolution showing beam-loading effects from driver energy spread. Top panels: On-axis accelerating field at injection (solid black) and at 17.0 mm (dashed black), shown with color-coded driver current profiles and shaded black witness current. Increased driver energy spread accelerates depletion, causing field flattening around the witness beam position. Bottom panels: Detailed view of the flattened accelerating field region across the witness bunch, demonstrating how driver depletion contributes to reduced energy spread in the witness beam. 
}
\label{fig: beam-loading-appendix}
\end{figure}

As discussed in the main text, certain latent effects which have an observable impact on the witness beam’s energy spread become noticeable at longer acceleration distances. Fig. \ref{fig: beam-loading-appendix} shows the underlying cause of the reduced energy spread behavior observed in the energy phase space diagrams in Fig. \ref{energyspread-stable}\ts(b). 
The top three panels show the longitudinal driver and witness current profiles at the capping point, and the longitudinal wakefield profiles generated by a driver beam of 5\%, 10\% and 50\% energy spread at injection (black solid line) and at the capping point (dashed line). The initial driver energy is 500 MeV and its charge 500 pC.   

While there is no difference between the wakefields at the beginning of the acceleration and the end for the low energy spread cases, the 50\% energy spread case shows that there is significant difference in wakefield amplitude at the witness position. 

As the driver beam depletes due to the self-created decelerating phase of the field, it causes the lowest energy electrons to eventually be recaptured at a phase of the wakefield that overlaps with the  witness beam. 
This additional, significant current at witness position provides local beam loading, and flattening of the longitudinal wakefield at this position. 

The bottom three panels are zoom-ins, highlighting the change of the wakefield due to this decelerated and re-accelerated charge. 
This flattened wakefield is responsible for the reduction in witness energy spread.

\section*{APPENDIX C: PARTICLE-IN-CELL SIMULATIONS}\label{Appendix: A-simualtions}
Simulations were performed using the quasi-3D Particle-in-Cell (PIC) code FBPIC. The drive beam was modelled as a normally distributed bi-Gaussian distribution with 1 million equally weighted macro-particles. The laser was approximated as a Gaussian pulse, solved at each time step. Its fields were applied to the macro-particles, enabling the recovery of ionization charge yields, which closely agreed with analytical ADK predictions. The background plasma (He$^{+}$)
was modelled as an electron-only species (16 macro particles particles per cell (PPC)), with convergence tests performed for all parameters before the production runs. Witness macro-particles were generated using the built in ADK module, with all fields incident on the helium ions kept safely below the ADK critical field ($\sim$400 GV/m), negating the need for a tunnelling and barrier suppression ionization tandem code. These particles represented He$^{2+}$ ions (32 PPC). 
A total of 3.24 million cells were used, with a cell size of 50 nm in the longitudinal direction (z) and 100 nm in the radial direction (r). This grid resolution allowed for accurate modelling of all key physical effects while considering the available simulation resources. All simulations were performed using a moving window and utilized the same box accompanied by 2 azimuthal modes with open box boundaries and default matched absorption layers applied transversely to the the ($135 \,\mu$m by $120 \,\mu$m) simulation box. 
A Galilean transformation of the co-moving grid was used to suppress the numerical Cherenkov instability and its spurious field emissions to prevent artificial degradation of the witness. 
Cross deposition current correction and an infinite stencil were used in all simulations, where the latter allowed for a more accurate solving of the dispersion relation compared to when using multi-MPI processes. A cosine flat-top cylindrically symmetrical plasma channel was used in all simulations, with a 2$\lambda_p$ entrance ramp implemented. Injection occurred in all cases with a fixed laser focal position of 0.5 mm in the lab frame, with the delay expressed in the main text behind the driver for the linked co-moving coordinate.
Fig. \ref{spat-temp} used test particles, which were spaced every \( dz \) and \( dy = dx = dr \) in Cartesian space. The center of the distribution was placed at the laser focus position, and the particles were instantaneously loaded at the appropriate simulation time step from pre-calculated arrays of momentum (all zero) and particle positions. To prevent these particles from influencing the simulation fields, their weight and subsequent current deposition were set to zero in this simulation set only.
Simulation post-processing used the same set of spatial and energy macroparticle cuts for laser injection runs.
\bibliography{hybridsBibtextidied2}
\end{document}